\documentclass[10pt,conference,twocolumn]{IEEEtran}

\usepackage{filecontents}
\usepackage[noadjust]{cite}
\usepackage{caption}
\usepackage{amsfonts}
\usepackage[dvips]{color, graphicx}
\usepackage[cmex10]{amsmath}
\usepackage{pgfplotstable}
\usepackage{array}
\usepackage{booktabs}
\usepackage{times}
\usepackage[normalem]{ulem} 

\ifCLASSOPTIONcompsoc
\usepackage[caption=false,font=normalsize,labelfon
t=sf,textfont=sf]{subfig}
\else
\usepackage[caption=false,font=footnotesize]{subfig}
\fi

\usepackage{flushend}
\usepackage{amsthm}
\usepackage{amsmath,amssymb}

\usepackage{algorithm}
\usepackage{algorithmic}

\usepackage{slashbox,multirow}

\addcontentsline{toc}{section}{Acknowledgment}

\long\def\symbolfootnote[#1]#2{\begingroup%
\def\thefootnote{\fnsymbol{footnote}}\footnote[#1]{#2}\endgroup}

\begin{document}
\title{Efficient Construction of Polar Codes}
\author{\IEEEauthorblockN{Wei Wang\IEEEauthorrefmark{1},
Liping Li\IEEEauthorrefmark{1}}
\IEEEauthorblockA{\IEEEauthorrefmark{1}Key Laboratory of Intelligent Computing and Signal Processing of the Ministry\\
of Education of China Anhui University, China,\\
Emails:  wangwei\_ahu@yahoo.com, liping\_li@ahu.edu.cn}}

\maketitle
\begin{abstract}
The construction of polar codes for channels other than BECs requires
sorting of all bit channels and then selecting
the best $K$ of them for a block length $N=2^n$.
In this paper, two types of partial orders (PO) of
polar codes are incorporated in the
construction process to decrease the required computations. Three sets,
corresponding to the good bit channels ($\mathcal{I}$),
the frozen bit channels ($\mathcal{F}$),
and the undetermined bit channels ($\mathcal{U}$), are selected by applying PO relations.
The POs are channel independent and are therefore universal for all binary-input discrete memoryless channels. For a given specific channel, a new process, called Dimension Reduction (DR), is proposed in this paper to further
reduce the size of $\mathcal{U}$.
Our studies show that for $N=10$ and the
code rate $R=0.5$ (being the worst code rate), incorporating PO relations alone can determine 50\%
of the bit channels ($|\mathcal{I}| + |\mathcal{F}| \approx N/2$).
With our proposed DR, this number of the determined
bit channels goes up to 82\%, which brings a
significant reduction of computations in the construction of polar codes.
\end{abstract}
\begin{IEEEkeywords}
Polar codes, polar code construction, partial orders, dimension reduction
\end{IEEEkeywords}
\section{Introduction}\label{sec_ref}

Polar codes have attracted researchers from around the world since its introduction in \cite{arikan_iti09}. It's proven in \cite{arikan_iti09} that polar codes can achieve the channel capacity at a low encoding and decoding complexity of $\mathcal{O}(N \log N)$ for binary-input discrete memoryless channels (B-DMC). A successive cancellation (SC) decoder was proposed to decode $N$ synthesized channels $W_N^{(i)}$, $i=1, 2,\cdots , N$. For a polar code with a code rate $R$, the $K=\lfloor NR \rfloor$ best bit channels $W_N^{(i)}$ are used to transmit information bits.
The quality of a channel $W_N^{(i)}$ can be measured, for example, by the error probability $P_e(W_{N}^{(i)})$. However, determining the quality of the channels is in general not an easy task except for BEC channels.

Let $\mathcal{I}$ denote the set containing $K$ indices of the good bit channels and
$\mathcal{F}$  the set containing the frozen bit channels' indices.
Let $W: \mathcal{X} \rightarrow \mathcal{Y}$ be the underlying channel with a transition probability $W(y|x)$.
As stated previously, sorting the bit channels are difficult because of the huge number of the output alphabets
$|\mathcal{Y}|^N$. 
Monte-Carlo simulations are proposed in \cite{arikan_iti09}
to sort the bit channels which has a high complexity of $\mathcal{O}(M N \log N)$ ($M$ being the iterations
of the Monte-Carlo simulations). Density evolutions are used in \cite{mori_isit09}\cite{mori_icl09}
to construct polar codes. The density evolution process involves function convolutions whose precisions are
limited by the complexity. In \cite{vardy_it13}, bit channel approximations are proposed with a controlled complexity
of $\mathcal{O}(N \cdot \mu^2 \log \mu)$ ($\mu$ being a user defined
parameter to limit the number of output alphabet at each approximation stage).
Another family of polar code construction is to use Gaussian Approximations (GA) on AWGN channels
\cite{trifonov_itc12,wu_icl14,niu_16}. The GA method is inherently limit by
the approximation function \cite{niu_16} and is therefore not clear whether GA is applicable
to all block lengths.

The partial order (PO) relations are reported in  \cite{mori_isit09}\cite{schurch_isit16}.
We borrow the notations from \cite{schurch_isit16} to denote
the two types of POs: $\preceq_1$ and $\preceq_2$. The first type of PO, $\preceq_1$, is studied
in \cite{schurch_isit16} which orders the bit channels with the same Hamming weight (here the Hamming weight
being that of the binary expansions of the bit channel indices). The second type of PO, $\preceq_2$, are
stated in \cite{mori_isit09}. The PO $\preceq_2$ orders the channels with different Hamming weight. These two POs
are channel independent: the orders are universal for all of underlying channels. Note that the
POs we study in this paper are different from the PO of \cite{alsan_isit14} in which stochastic
degradation relations are studied between two binary-input discrete  memoryless channels.

The aforementioned constructions in \cite{mori_isit09, mori_icl09, vardy_it13, trifonov_itc12, wu_icl14,niu_16}
sort all of the $N$ bit channels. The inherent ordering of polar codes  from $\preceq_1$ and $\preceq_2$
is not utilized before sorting all the bit channels.  Therefore, it is not clear how much we can save in
the construction by applying POs first. In this paper, we try to apply $\preceq_1$ and $\preceq_2$
before sorting all $N$ bit channels and to quantify the savings in this process.
Three sets, corresponding to the good bit channels ($\mathcal{I}$),
the frozen bit channels ($\mathcal{F}$),
and the undetermined bit channels ($\mathcal{U}$), are selected by applying the two POs.
Any sorting algorithm, for example, the approximation in \cite{vardy_it13}, can be applied to sort the
bit channels in $\mathcal{U}$. The smaller the size of $\mathcal{U}$, the larger the savings are.
The POs are channel independent and are universal for all B-DMCs. Therefore, the results for a given block length $N$ and a code rate $R$ only need to be calculated once. For a given channel condition, to further reduce the size of $\mathcal{U}$, we introduce a Dimension Reduction (DR) method.
DR works by first sorting the channels at a block length $N' = 2^{n'} < N = 2^n$.
The resulted ordering of $N'$ channels is used
together with $\preceq_1$ and $\preceq_2$ again to move the channels in  $\mathcal{U}$ to either $\mathcal{I}$
or $\mathcal{F}$. The calculations in our study show that for $N=10$ and the
code rate $R=0.5$ (being the worst code rate), incorporating PO relations alone can determine 50\%
of the bit channels ($|\mathcal{I}| + |\mathcal{F}| \approx N/2$). With our proposed DR, the determined portion of the channels
goes up to 82\%, which brings a
significant reduction in the construction of polar codes.

The notations in this paper are as the following.
The notation $v_1^N$ is used to represent a row vector with elements $(v_1,v_2,...,v_N)$.
The $n$-bit binary expansion of an integer $i$ is written as $i=(i_{n},i_{n-1},...,i_1)_b$.
Given a vector $v_1^N$, the
vector $v_i^j$ is a subvector $(v_i, ..., v_j)$ with $1 \le i,j \le N$. If there is a set $\mathcal{A} \in \{1,2,...,N\}$,
then $v_{\mathcal{A}}$ denotes a subvector with elements in $\{v_i, i \in \mathcal{A}\}$.

The rest of the paper is organized as the following. Section \ref{sec_polar_partial} introduces the basics of polar codes
and the partial orders we used in this paper.  The construction using the two partial orders
is presented in Section \ref{sec_po_dr}. Also presented in Section \ref{sec_po_dr} is
our proposed dimension reduction method. The efficiency of the proposed construction is given in Section \ref{sec_numerical}.
Conclusion remarks are presented at the end in Section \ref{sec_con}.

\section{Polar Code and Partial Orders} \label{sec_polar_partial}
The first part of this section includes the relevant basics of polar codes from \cite{arikan_iti09}.
For all other details of polar codes, please refer to \cite{arikan_iti09}.
The second part is to restate some results of partial orders from \cite{mori_isit09} and \cite{schurch_isit16}.

\subsection{Polar Codes}
Let $W$ be any binary-input discrete memoryless channel (B-DMC) with a transition probability $W(y|x)$.
The input alphabet $\mathcal{X}$ takes values in $\{0,1\}$ and
the output alphabet is $\mathcal{Y}$. The generator matrix $G = B F^{\otimes n}$
where $B$ is a permutation matrix and
$F=\left[\begin{smallmatrix} 1&0 \\ 1&1 \end{smallmatrix}\right]$. The operation $F^{\otimes n}$
is the $n$th Kronecker power of $F$ over the binary field $\mathbb{F}_2$.
Polar codes synthesize $N=2^n (n \ge 1)$ bit channels, $\{{W_N^{(i)}}\}_{i=1}^N$,
out of $N$ independent copies of $W$. The transformation has a tree structure in \cite{arikan_iti09},
which has
a basic one-step channel transformation defined as $(W,W) \mapsto (W^{'},W^{''})$, where
\begin{eqnarray}\nonumber
W^{'}(y_1,y_2|u_1) &=& \sum_{u_2}\frac{1}{2}W(y_1|u_1\oplus u_2)W(y_2|u_2) \\ \label{eq_wp}
 \\ \label{eq_wpp}
W^{''}(y_1,y_2,u_1|u_2) &=& \frac{1}{2}W(y_1|u_1\oplus u_2)W(y_2|u_2)
\end{eqnarray}
A bit channel $i-1=(i_{n}, i_{n-1}, ...,i_1)_b$ ($1 \le i \le N$) is transformed in
each tree level according to
the bit of $i$ at that level: at tree level $ 1 \le k \le n$, bit channel $i$ takes $W^{'}$
if $i_k = 0$. Otherwise $W^{''}$ is taken at level $k$.
The Bhattacharyya parameters of $W^{'}$ and $W^{''}$ satisfy the following conditions:
\begin{eqnarray} \label{eq_zp}
Z(W^{''}) = Z(W)^2 \\ \label{eq_zpp1}
Z(W^{'}) \le 2Z(W)-Z(W)^2 \\ \label{eq_zpp2}
Z(W^{'}) \ge Z(W) \ge Z(W^{''})
\end{eqnarray}
Note that for binary
erasure channels (BEC), the Bhattacharyya parameter $Z(W^{'})$ has an exact expression $Z(W^{'}) = 2Z(W)-Z(W)^2$,
resulting in a recursive calculation of the Bhattacharyya parameters of the final bit channels.
Finally, after the channel transformations, the transition probability for bit channel $i$ is defined as

\begin{eqnarray}\label{eq_wn_slpit}
W_N^{(i)}(y_1^N,u_1^{i-1}|u_i) = \sum_{u_{i+1}^{N} \in \mathcal{X}^{N-i}}\frac{1}{2^{N-1}}W^N(y_1^N|u_1^NG)
\end{eqnarray}
where $W^N(\cdot)$ is the underlying vector channel ($N$ copies of the channel $W$).

\subsection{Partial Orders}
As in \cite{vardy_it13}\cite{schurch_isit16}, we write
$W_N^{(j)} \preceq W_N^{(i)}$ if bit channel $j$ is stochastically degraded with
respect to bit channel $i$ ($1 \le i,j \le N$).

To introduce the first PO, $\preceq_1$, some of the notations and theorems are restated from \cite{schurch_isit16}
without providing proofs. Denote $\mathcal{Z}_n = \{1,2,...,n\}$. The following statement defines
a relation between two numbers with the same Hamming weight \cite{schurch_isit16}.
\newtheorem{definition}{Definition}
\begin{definition}\label{definition_1}
For $1 \le i,j \le N$, let $i-1 = (i_{n-1},i_{n-2},...,i_1)_b$ and $j-1 = (j_{n-1},j_{n-2},...,j_1)_b$ be
the binary expansion of $i-1$ and $j-1$ respectively.
We write $j \nearrow i$ if there exist $l$, $l^{'} \in \mathcal{Z}_n$ with $l<l^{'}$ such that
\begin{enumerate}
\item $j_l=1$ and $j_{l'}=0$
\item $i_l=0$ and $i_{l^{'}}=1$
\item For all $k \in \mathcal{Z}_n\backslash \{l,l^{'}\}: j_{k}=i_{k}$
\end{enumerate}
\end{definition}
In Definition \ref{definition_1}, $j-1$  can be obtained from $i-1$ by switching a
higher position 1 (at position $l'$) with a lower position 0 (at position $l$). The
two numbers with such a relation is denoted as $j \nearrow i$.
The following theorem \cite{schurch_isit16} states the relationship between two bit channels
$W_N^{(i)}$ and $W_N^{(j)}$ with $j \nearrow i$.
\newtheorem{theorem}{Theorem}
\begin{theorem}\label{theorem_1}
If $j \nearrow i$ then $W_N^{(j)}$ is stochastically degraded with respect to $W_N^{(i)}$.
\end{theorem}

With Theorem \ref{theorem_1}, bit channels with the same Hamming weight can be ordered. The order obtained from Theorem \ref{theorem_1} is written as $\preceq_1$, which is equivalent to $\preceq$.
The second partial order comes from \cite{mori_isit09} and is restated below in Theorem \ref{theorem_2}.
\begin{theorem}\label{theorem_2}
For $1 \le i, j \le N$, $W_N^{(j)} \preceq_2 W_N^{(i)}$ if and only if
$i_t = 1$ when $j_t = 1$ for any $t \in \mathcal{Z}_n$, where $(i_{n},i_{n-1},...,i_1)$ and
$(j_{n},j_{n-1},...,j_1)$ are binary expansions of $i-1$ and $j-1$ respectively.
\end{theorem}

With a symmetric underlying channel $W$, it's proven \cite{mori_isit09} that
$W_N^{(j)} \preceq_2 W_N^{(i)} \implies W_N^{(j)} \preceq W_N^{(i)}$. Therefore,
the second partial order $\preceq_2$ can also be used to order the bit channels.

To the authors' knowledge, the work in \cite{schurch_isit16} is the first to combine the two partial orders to speed up the construction of polar codes. There is a general algorithm in \cite{schurch_isit16} where no further details are provided. Here in this paper, we provide detailed algorithms and specific results to quantify the advantage of the POs. A Dimension Reduction  is also proposed to further reduce the computations needed in the construction stage.

\section{Construction Using POs and Dimension Reduction}\label{sec_po_dr}

\subsection{Construction with POs} \label{sec_po}
As noted in \cite{schurch_isit16}, partial orders can only determine the relationship of some of the bit channels.
If we want to sort all $N$ bit channels, it necessary to obtain all the $(
    \begin{smallmatrix}
      N \\
      2
    \end{smallmatrix})$ ($N$ choose 2) relationships.
It's not hard to see that we still have to use the sorting algorithms as in \cite{vardy_it13}
to obtain the error probability  $\{P_e(W_N^{(i)})\}_{i=1}^{N}$ for all the bit channels except for a few best channels.

However, in the polar code design, what is of interest is to find $K=\lfloor NR \rfloor$
best channels among $N$ bit channels. In this regard, the two partial orders, $\preceq_1$ and $\preceq_2$,
can be used to pre-determine the elements of  $\mathcal{I}$, $\mathcal{F}$, and $\mathcal{U}$
before any sorting algorithm is applied. The sets $\mathcal{I}$, $\mathcal{F}$, and $\mathcal{U}$
store good channel indices, frozen channel indices, and undetermined channel indices, respectively.
Applying $\preceq_1$ and $\preceq_2$,
some of the relationships of the $N$ bit channels can be
determined. The following proposition can be used to determine whether a bit channel should be in $\mathcal{I}$,
$\mathcal{F}$, or $\mathcal{U}$.
\newtheorem{proposition}{Proposition}
\begin{proposition} \label{proposition_1}
If a bit channel $i$ is better than at least $N-K$ bit channels, then $i \in \mathcal{I}$.
If a bit channel $j$ is worse than at least $K$ bit channels, then $j \in \mathcal{F}$.
All other bit channels are in the undetermined set $\mathcal{U}$.
\end{proposition}
The proof of this proposition is quite straightforward and is therefore omitted here.

As in \cite{schurch_isit16}, the two POs can be combined to obtain more relations
among bit channels. An example can show the power of the combinations of $\preceq_1$ and $\preceq_2$.
Let $N= 2^5 = 32$. The relationship between bit channel $i-1=(10110)_b$ and $j-1=(00101)_b$ can't be
determined purely from $\preceq_1$ or $\preceq_2$. The Hamming weights of $i$ and $j$ are not the same.
Therefore $\preceq_1$ can't be used to order them. The PO $\preceq_2$ can't be
applied either because the first position does not satisfy Thoerem \ref{theorem_2}: $i_1=0$ but $j_1 = 1$.
However, bit channel $i'-1=(00110)_b$ and $j-1=(00101)_b$ can be determined from Theorem
\ref{theorem_1} as $W_{32}^{(j)} \preceq_1 W_{32}^{(i')}$. Then bit channel $i'$ and $i$ can be
determined from theorem \ref{theorem_2} as: $W_{32}^{(i')} \preceq_2 W_{32}^{(i)}$. From
the transitivity of the POs,  we have $W_{32}^{(j)} \preceq_1 W_{32}^{(i')} \preceq_2 W_{32}^{(i)}$
which is equivalent to $W_{32}^{(j)} \preceq W_{32}^{(i')} \preceq W_{32}^{(i)}$.


In our algorithm, $R_p$ is defined as a $N \times N$ matrix whose $i$th row  stores the relationship between bit channel $i$ and bit channel $j$ ($i>j$). Only the lower triangular part (excluding the diagonal) of the matrix $R_p$ stores the relationship between all bit channels. If $W_N^{(i)} \succ W_N^{(j)}$, then $R_p(i,j) = 1$. If $W_N^{(i)} \prec W_N^{(j)}$, then $R_p(i,j) = -1$. If the relationship between $W_N^{(i)}$ and $W_N^{(j)}$ is unknown, then $R_p(i,j) = 0$. The rest of $R_p$ can be arbitrary.
In practical implementations, the undefined part of $R_p$ can be saved. In Algorithm \ref{algorithm_po}, there is no $-1$s in $R_p$. This is because for any bit channel $j$ ($j < i$), it must be worse than bit channel $i$ purely from the two POs.

\begin{algorithm}
\caption{Function $R_p$ = PO\_relation(n).
Find out all the relationships that POs can determine, and store the results in a matrix $R_p$ }
\label{algorithm_po}
\begin{algorithmic}[1] 
\REQUIRE ~~\\ 
$n$;\\
\ENSURE ~~\\ 
a matrix $R_p$;\\
\STATE $N$=$2^n$ \;
\STATE $ R_p = zeros(N,N) $ \;
\FOR {$i = 2$ to $N$}
\STATE $(i_{n},i_{n-1},...,i_1)_b \rightarrow a$\;
    \FOR {$j = 1$ to $i-1$}
\STATE $(j_{n},j_{n-1},...,j_1)_b \rightarrow b$\;
\STATE  $x = a-b$\;
                \IF {there are at least the same positions of 1s' which are larger than the positions of -1s}
\STATE            $R_p(i,j) = 1$\;
                \ENDIF
    \ENDFOR
\ENDFOR
\RETURN $R_p$; 
\end{algorithmic}
\end{algorithm}

\begin{figure}
{\par\centering
\resizebox*{3.0in}{!}{\includegraphics{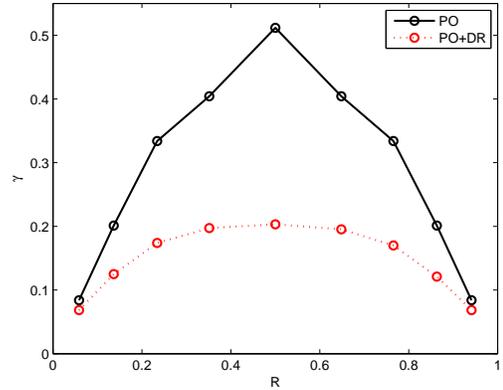}} \par}
\caption{
The value of $\gamma$ as a function of the code rate $R$ with POs and PO+DR. The underlying channel is AWGN
with a SNR of 1 dB. The block length is $N=2^9$.}
\label{fig_vbr}
\end{figure}

Once the relationship matrix $R_p$ is obtained, we can obtain two vectors $v_1^N$ and $w_1^N$: the entry $v_i$ of  $v_1^N$ corresponds to the number of bit channels that are worse than bit channel $i$, and the entry $w_j$ of $w_1^N$ corresponds
to the number of bit channels that are better than bit channel $j$. Algorithm \ref{algorithm_po} shows
how to obtain the matrix $R_p$. After obtaining $R_p$, process it to get the vectors $v_1^N$ and $w_1^N$. Then Proposition
\ref{proposition_1} can be applied to assign elements to the sets $\mathcal{I}$, $\mathcal{F}$, and $\mathcal{U}$.
What determines the remaining calculations of the sorting algorithms  is the size of $\mathcal{U}$.
For example, $|\mathcal{U}|$ determines the number of callings of the Tal-Vardy algorithm \cite{vardy_it13}.
In Fig.~\ref{fig_vbr}, the ratio $\gamma = |\mathcal{U}| / N$ is plotted for
$N=2^9$ (the solid circled line) which shows that the combinations of the two POs can determine 91\% of the
bit channels with $\gamma = 9\%$ for $R=0.1$.
To obtain a more clear picture of the bit channels whose relationships
are being determined and undetermined, a plot is shown
in Fig.~\ref{fig_relation_9}-(a) for $N=2^9$ and $R=0.5$. The black dots in Fig.~\ref{fig_relation_9}-(a)
correspond to value 1 and the white dots are value 0. From Fig.~\ref{fig_relation_9}-(a),
it can be seen that there are still rooms of white space which contribute to the
size of the undertermined set $\mathcal{U}$. In the following
subsection, a dimension reduction method is applied to further reduce the size of $\mathcal{U}$.

\subsection{Dimension Reduction}
\begin{figure}
\centering
\subfloat[Determined only from the two POs]{\includegraphics[width=1.7in]{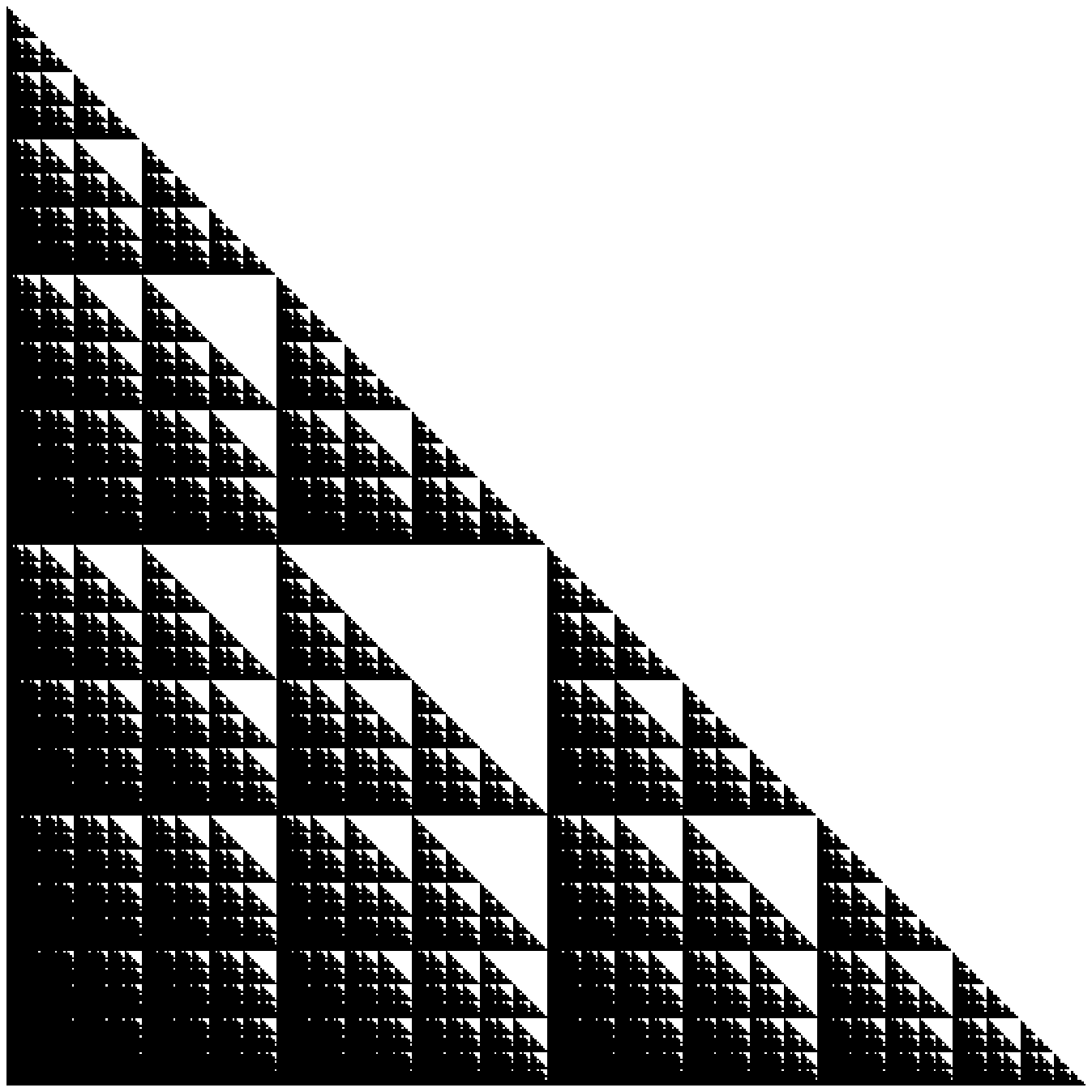}}
\hfil
\subfloat[Determined from the two POs and the dimension reduction]{\includegraphics[width=1.7in]{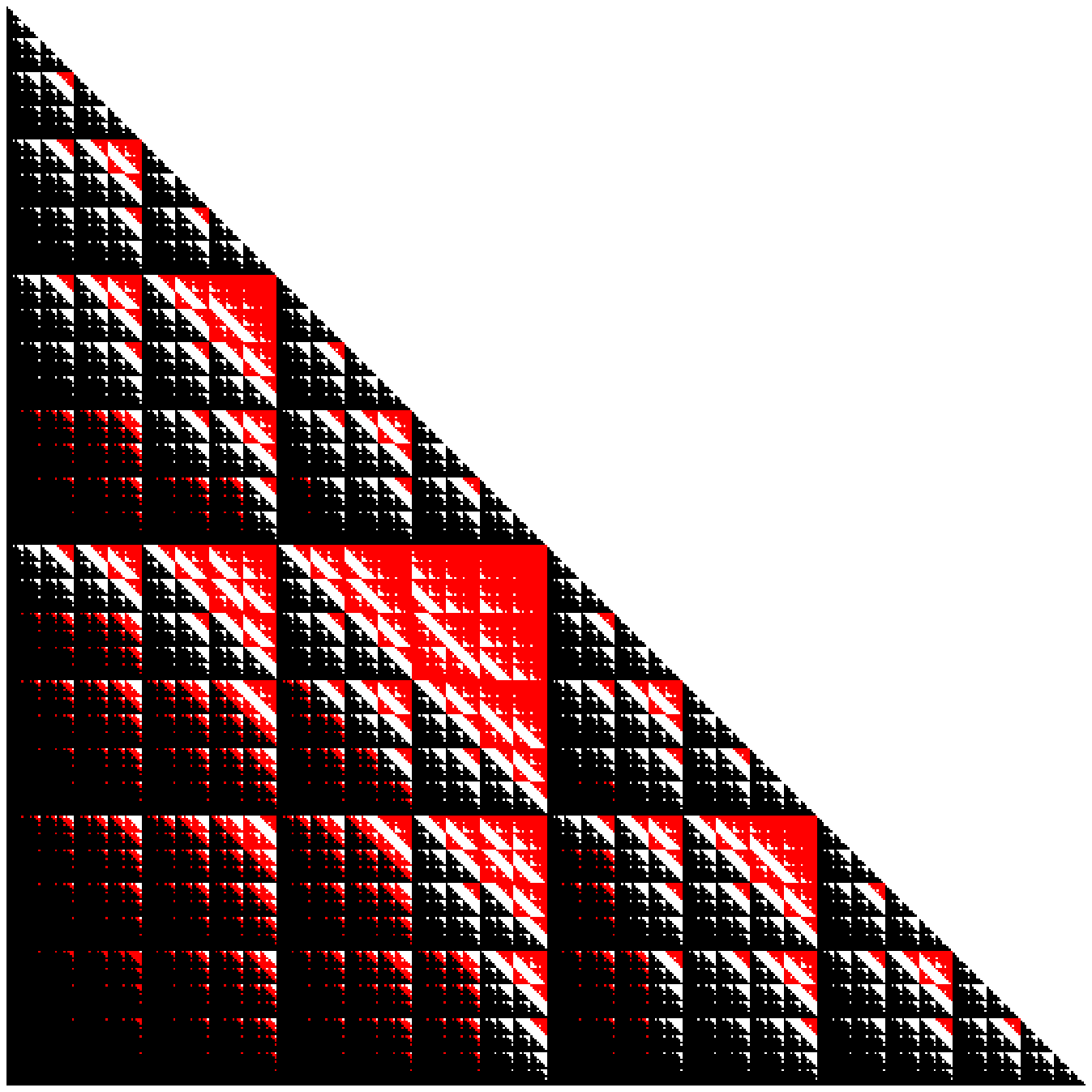}}
\caption{The relationship of all bit channels with $N=2^9$.
(a): Black dots indicate that the relationship is determined from the two POs
while white dots mean that the relationship can't be determined from the two POs.
(b):
The underlying channel is AWGN with a SNR of 1 dB. On top of the black dots, the red dots indicate
the additional relationships obtained from the dimension reduction.}
\label{fig_relation_9}
\end{figure}

The white space in Fig.~\ref{fig_relation_9}-(a) is caused by the fact that the relationship between some bit channels
can't be determined from POs $\preceq_1$, $\preceq_2$, or their combinations. Here is an
example. For $N=2^8$, consider bit channel $i-1=(10011110)_b$ and $j-1=(01101011)_b$. Applying the two POs
and their combinations yields no decision on whether $W_N^{(i)}$ is better or worse than $W_N^{(j)}$.

Let's take a closer look at the two bit channels in the example above.
Let $\mathcal{Z}_u$ denote the upper part of the descending set $\mathcal{Z}_n$
with $n_u = |\mathcal{Z}_u|$. Similarly $\mathcal{Z}_l$ is the lower part of
$\mathcal{Z}_n$ with $n_l = |\mathcal{Z}_l|$. The sizes satisfy $ 0 \le n_u, n_l \le n$ and $n_u+n_l=n$.
Divide the binary expansion of $i$ into
two consecutive parts: $i-1=(i_{\mathcal{Z}_u}, i_{\mathcal{Z}_l})_b$.
The upper part $i_{\mathcal{Z}_u}$ defines a new bit channel at the block length
$N_u = 2^{n_u} \le N=2^n$: $W_{N_u}^{(i_u)}$.
For the lower part, there is similarly a new bit channel.
For this specific example, $n_u=5$ and $n_l = 3$. Among the two new bit channels,
one of them can be determined: $W_{N_l}^{j_l} \preceq_2 W_{N_l}^{i_l}$. Only the relationship between the upper channels
can't be determined.

On the other hand, if the upper bit channels can be found to be  $W_{N_u}^{j_u} \preceq W_{N_u}^{i_u}$
through any sorting algorithm, then we can reach the conclusion that $W_N^{(j)} \preceq W_N^{(i)}$
from the following proposition.
\begin{proposition}\label{proposition_2}
Consider bit channel $i$ and $j$ $(1 \le i,j \le N)$
with $i-1=(i_n,i_{n-1},...,i_1)_b$ and $j-1=(j_n,j_{n-1},...,j_1)_b$. Divide the binary
expansions of $i-1$ and $j-1$ into two parts:  $i-1=(i_{\mathcal{Z}_u}, i_{\mathcal{Z}_l})_b$
and $j-1=(j_{\mathcal{Z}_u}, j_{\mathcal{Z}_l})_b$.
If the relationship between $W_{N_u}^{i_u}$ and $W_{N_u}^{j_u}$ is obtained for a given underlying channel $W$ and the relationship between $W_{N_l}^{i_l}$ and $W_{N_l}^{j_l}$ is channel independent, then

$W_{N_u}^{(i_u)} \succeq W_{N_u}^{(j_u)}$ and $W_{N_l}^{(i_l)} \succeq W_{N_l}^{(j_l)} \Rightarrow W_N^{(i)} \succeq W_N^{(j)}$
\end{proposition}
The proof of Proposition \ref{proposition_2} is omitted in this paper due to the space limit.

In this paper, a Dimension Reduction (DR) method is proposed to reduce the
size of the undetermined set $\mathcal{U}$ (also to fill in the white space in
Fig.~\ref{fig_relation_9}-(a)). This DR is based on Proposition \ref{proposition_2}
which needs to call a sorting algorithm to order the bit channels at a upper block
length $N_u = 2^{n_u} < N=2^n$ (also where the name DR coming from).
A parameter to be determined is the upper level $n_u$. In this paper, we take
$n_u = n-3$ empirically. At this lower level, a relationship matrix, $R_u$ is first
obtained from Algorithm \ref{algorithm_po}. Then the undetermined part of $R_u$ is
filled in by the order obtained from calling any sorting algorithm,
for example \cite{mori_isit09} \cite{vardy_it13}. The algorithm
of this part is not provided in the paper due to the space limit.
Then, the original relationship matrix
$R_p$ is updated by DR. This part is provided in Algorithm \ref{algorithm_dr}. Finally, a top-level
Algorithm is provided in Algorithm \ref{algorithm_top}. The results of the proposed
DR can be seen from the red dots in Fig.~\ref{fig_relation_9}-(b), which show the
additional relationships obtained from the DR.

\begin{algorithm}
\caption{Function $R_p$ = DR\_relation($R_p$,$R_u$). Apply the DR to determine
more relationships among bit channels.}
\label{algorithm_dr}
\begin{algorithmic}[1] 
\REQUIRE ~~\\ 
$R_p$, $R_{u}$;\\
\ENSURE ~~\\ 
$R_p$;\\
\STATE $//$ We divide a channel into two parts,eg:
\STATE $//$ ${\underbrace{xx\cdots x}_{n_u}\underbrace{yy\cdots y}_{n_l}}$\;
\STATE $//$ $n_u+n_l=n$\;
\STATE $N_l=2^{n_l}$, $N_u=2^{n_u}$\;
\STATE $i_u, j_u \in [1,N_u]$ $//$ two channels from the upper part\;
\STATE $i_l, j_l \in [1,N_l]$ $//$ two channels from the lower part\;\;
\STATE $//$ Try to determine the relationship of the following two channels
\STATE $//$ channel $i$£º\;
\STATE $i=(i_u-1)\times N_l+i_l$\;
\STATE $//$ channel $j$£º\;
\STATE $j=(j_u-1)\times N_l+j_l$\;
\STATE $//$  the relationship between $i_u$ and $j_u$ can not be obtained from POs (it's obtained from a sorting algorithm) and the relationship between $i_l$ and $j_l$ can be obtained from POs.
\IF {($i_u$ is better than $j_u$) and ($i_l$ is not worse than $j_l$)}
    \IF {$i>j$}
    \STATE $R_p(i,j)=1$ \;
    \ELSE
\STATE $//$ bit channel $j$ is worse than bit channel $i$
    \STATE $R_p(j,i)=-1$ \;
    \ENDIF
\ENDIF
\RETURN $R_p$; 
\end{algorithmic}
\end{algorithm}

\begin{algorithm}
\caption{Determine elements of $\mathcal{I}$, $\mathcal{F}$, $\mathcal{U}$.}
\label{algorithm_top}
\begin{algorithmic}[1] 
\REQUIRE ~~\\ 
$n$, $R$, $snr$;\\
\ENSURE ~~\\ 
$\mathcal{I}$, $\mathcal{F}$, $\mathcal{U}$;\\
\STATE $N=2^n$, $K=\lfloor N\times R \rfloor$, $N_f=N-K$\;
\STATE $n_u=n-3$, $N_u=2^{n_u}$\;
\STATE  $R_p=$ PO\_relation(n). $//$ Call Algorithm \ref{algorithm_po}, return the matrix $R_p$ on the basis of $n$.\;
\STATE $//$ Assume $R_u$ at the low dimension is already calculated at the input $snr$\;
\STATE $R_p$ = DR\_relation($R_p$, $R_u$) $//$ Call Algorithm \ref{algorithm_dr}, return the matrix $R_p$.\;
\STATE $//$ vector $s_1^N$: $s_i$ the number of bit channels worse than bit channel $i$ \;
\STATE $//$ vector $f_1^N$: $f_i$ the number of bit channels better than bit channel $i$ \;
\STATE $//$ this function counts the number of channels worse than each bit channel and better
than each bit channel \;
\STATE ($s_1^N$, $f_1^N$) = counting\_channels($R_p$)\;
\STATE $\mathcal{I}=find(s_1^N \geq  N_f)$\;
\STATE $\mathcal{F}=find(f_1^N \geq K)$\;
\STATE Put elements not in $\mathcal{I}$ and $\mathcal{F}$ to $\mathcal{U}$ \;
\end{algorithmic}
\end{algorithm}




\section{Numerical Results}\label{sec_numerical}
In this section, the efficiency of applying POs and DR is quantified. The underlying
channel is the AWGN with a SNR of 1 dB. When applying the DR, the algorithm called to sort the lower dimension channels is from \cite{vardy_it13}. For ease of description, we use PO to refer to the results applying the two POs and PO + DR meaning both POs and DR are applied. All our construction results in this section match the constructions in [4]. Therefore, in the following, we only show  the efficiency results.

Fig.~\ref{fig_vbr} shows the percentage of the remaining
bit channels which can't be determined from POs, or from PO+DR.
This percentage is defined
as $\gamma$ in Section \ref{sec_po}. The circled line
is the value of $\gamma$ as a function of the code rate $R$ for the block length $N=2^9$.
The line with stars in Fig.~\ref{fig_vbr} is $\gamma$ after applying POs and DR. It can
be seen from Fig.~\ref{fig_vbr} that the value of $\gamma$ decreases with the increase of $R$ when $R > 0.5$ and
increases with $R$ when $R < 0.5$. At the code rate
$R$ = 0.5, both PO and PO + DR have the largest value of $\gamma$.
But the savings of applying DR is also the largest at $R=0.5$.  For a small code rate $R \le 0.1$, less than 10\% of the remaining channels are left in the undetermined set $\mathcal{U}$.
Both PO and PO+DR have the largest value of $\gamma$ at $R=0.5$.

\begin{figure}
{\par\centering
\resizebox*{3.0in}{!}{\includegraphics{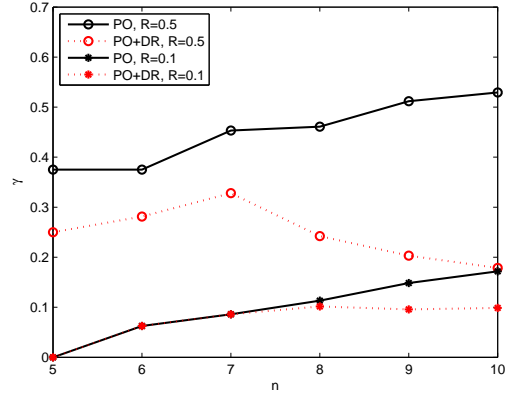}} \par}
\caption{
 The value of $\gamma$ with two different code rates:  $R=0.1$ and $R=0.5$. The underlying channel is AWGN
 with a SNR of 1 dB.}
\label{fig_vcl}
\end{figure}

Fig.~\ref{fig_vcl} shows the relationship between $\gamma$ and the block length for two
fixed code rates: $R=0.5$ and $R=0.1$. The two circled lines are for $R=0.5$: the solid line
is with POs and the dashed one is PO+DR. Similarly the two lines with stars are values of $\gamma$ for $R=0.1$.
From Fig.~\ref{fig_vcl}, we can see that the number of undetermined channels slowly increases with the
block length when applying POs. The same is true for PO+DR although there is a very small variation.
From Fig.~\ref{fig_vcl}, it can be seen that PO+DR can bring around 82\% of the savings
in terms of the computation time for polar code construction with a code rate $R=0.5$ and $N=1024$.

%
%

\section{Conclusion}\label{sec_con}
In this paper, we apply the partial orders of polar codes to decrease the computation complexity in the construction.
We show clearly that POs can indeed bring big savings in constructing polar codes. The remaining undetermined bit
channels after applying POs are further processed by the proposed dimension reduction and significant savings are achieved.

\section*{Acknowledgment}
This work was supported in part by National Natural Science Foundation of China through grant 61501002, in part by Natural Science Project of Ministry of Education of Anhui through grant KJ2015A102,  in part by the Key Laboratory Project of the Key Laboratory of Intelligent Computing and Signal Processing of the Ministry of Education of China, Anhui University, in part by Talents Recruitment Program of Anhui University.
\bibliography{../ref_polar}
\bibliographystyle{IEEEtran}
\end{document}